\PassOptionsToPackage{numbers}{natbib}
\documentclass{article}

\usepackage[preprint]{neurips_2024}

\usepackage[utf8]{inputenc} 
\usepackage[T1]{fontenc}    
\usepackage{hyperref}       
\usepackage{url}            
\usepackage{booktabs}       
\usepackage{amsfonts}       
\usepackage{nicefrac}       
\usepackage{microtype}      
\usepackage{xcolor}         
\usepackage{fontawesome}
\usepackage{marvosym}
\hypersetup{pdfborder={0 0 0}}
\usepackage{graphicx}
\usepackage{pifont}
\usepackage{amsmath}
\usepackage{wrapfig}
\usepackage{subcaption}

\hypersetup{
    colorlinks=true,
    linkcolor=red,
    citecolor=cyan,
    filecolor=magenta,      
    urlcolor=magenta,
    }

\usepackage{tcolorbox}

\tcbuselibrary{breakable, listings}

\newtcblisting{promptbox}[1][]{
    title       = {System Prompt},
    fonttitle   = \bfseries,
    breakable,
    listing only,
    listing options = {
        basicstyle  = \ttfamily\scriptsize,
        breaklines  = true,
    },
    colback     = gray!5,
    colframe    = gray!50,
    boxrule     = 0.5pt,
    arc         = 3pt,
    #1,
}

\title{SWE-CI: Evaluating Agent Capabilities \\ in Maintaining Codebases via Continuous Integration}

\author{%
  Jialong Chen$^{1,3,*}$\href{mailto:chenjlong7@mail2.sysu.edu.cn}{\Letter},
  Xander Xu$^{2,}$\thanks{Equal contribution.}\,\,\,\href{mailto:yanquan.xx@alibaba-inc.com}{\Letter},
  Hu Wei$^{2}$\,\href{mailto:kongwang@alibaba-inc.com}{\Letter},
  Chuan Chen$^{1,\dagger}$\,\href{mailto:chenchuan@mail.sysu.edu.cn}{\Letter},
  Bing Zhao$^{2,}$\thanks{Corresponding author.}\,\,\,\href{mailto:xiongdao@alibaba-inc.com}{\Letter}%
  \medskip \\
  $^{1}$Sun Yat-sen University, $^{2}$Alibaba Group, $^{3}$Skylenage
  \smallskip \\[4pt]
  \raisebox{-0.3em}{\includegraphics[height=1.2em]{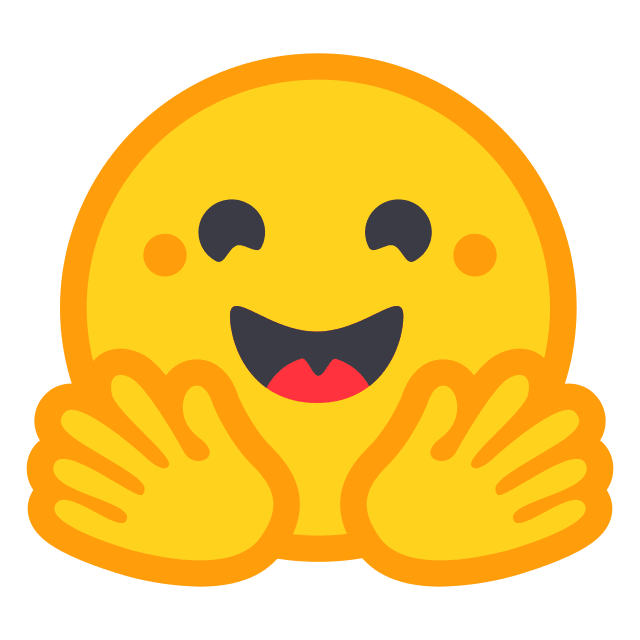}}\,\texttt{Hugging Face: }\href{https://huggingface.co/datasets/skylenage/SWE-CI}{\texttt{https://huggingface.co/datasets/skylenage/SWE-CI}}
  \\[4pt]
  \raisebox{-0.3em}{\includegraphics[height=1.2em]{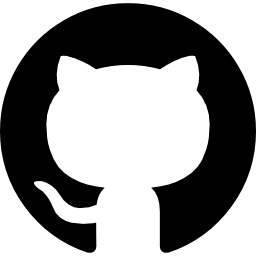}}\,\texttt{Github: }\href{https://github.com/SKYLENAGE-AI/SWE-CI}{\texttt{https://github.com/SKYLENAGE-AI/SWE-CI}}
}

\begin{document}

\maketitle

\begin{abstract}
Large language model (LLM)-powered agents have demonstrated strong capabilities in automating software engineering tasks such as static bug fixing. However, in the real world, the development of mature software is typically predicated on  complex requirement changes and long-term feature iterations --- a process that static, one-shot repair paradigms fail to capture. To bridge this gap, we propose \textbf{SWE-CI}, the first repository-level benchmark built upon the Continuous Integration loop, aiming to shift the evaluation paradigm for code generation from static, short-term \textit{functional correctness} toward dynamic, long-term \textit{maintainability}. The key insight is simple: \textbf{Maintainability can be revealed by tracking how functional correctness changes over time.} The benchmark comprises 100 tasks, each deriving from a real-world code repository with a development history spanning an average of 233 days and 71 consecutive commits. SWE-CI requires agents to systematically resolve these tasks through dozens of rounds of analysis and coding iterations. SWE-CI provides valuable insights into how well agents can sustain code quality throughout long-term evolution.
\end{abstract}

\section{Introduction}


Automating software engineering has long been a central objective in the field of artificial intelligence. In recent years, breakthroughs in large language models (LLMs) have lent substantial momentum into this pursuit --- from code completion and test generation to end-to-end program repair, LLM-driven agents have demonstrated capabilities rivaling those of human developers across multiple benchmarks. The concurrent evolution of coding benchmarks has played a pivotal role in this progress, providing both rigorous capability measurement and clear research direction.

At the code generation level, HumanEval \cite{1}, MBPP \cite{2} and LiveCodeBench \cite{3} established the single-file synthesis paradigm. At the repository level, SWE-bench \cite{4} introduced the ``Issue-to-PR'' paradigm, requiring models to generate patches within complete repository contexts. At the agent interaction level, Terminal-bench \cite{5} and $\tau$-bench \cite{6} further broadened the evaluation scope to encompass terminal operations and multi-turn tool-use. Together, these efforts have established a multi-granularity and multi-scenario evaluation ecosystem for code intelligence.

Despite the breadth and depth of this ecosystem, its underlying paradigm exhibits one fundamental limitation: Existing benchmarks almost exclusively focus on measuring the agent's ability to write functionally correct code. However, in the real world, successful software is rarely achieved overnight; it is the result of long-term maintenance. Lehman's Laws reveal that software quality inherently degrades as maintenance progresses \cite{7}, while classic literature has established that maintenance activities account for 60\% to 80\% of total software lifecycle costs \cite{8}. Therefore, there is an urgent need to design new benchmarks to effectively reflect the ability of models to maintain code.

The reason this capacity has long been absent from evaluation is rooted in the prevailing benchmark paradigm itself. From HumanEval and LiveCodeBench to SWE-bench and Terminal-Bench, existing benchmarks universally adopt a snapshot-style protocol: the agent receives a single, complete requirement and produces a one-shot solution. Under this paradigm, an agent that hard-codes a brittle fix and one that writes clean, extensible code may both pass the same test suite --- their difference in maintainability is simply invisible. It becomes visible only when the codebase must evolve: new requirements arrive, interfaces change, and modules must be extended. At that point, the cost of earlier design decisions compounds, and an agent that routinely produces poorly structured code will find each successive modification harder, eventually failing to keep pace. This yields a critical insight: \textbf{an agent's ability to maintain code can only be revealed through long-term evolution, where the consequences of past decisions accumulate over successive changes.}

Building upon this insight, we propose \textbf{SWE-CI} (\textbf{S}oft\textbf{W}are \textbf{E}ngineering -- \textbf{C}ontinuous \textbf{I}ntegration), a novel benchmark designed to evaluate how well agents maintain code across long-term code evolution. SWE-CI comprises 100 tasks, each defined by a \textit{base commit} and a \textit{target commit} from a real-world repository, with an average of 233 days and 71 consecutive commits of authentic evolutionary history in between. SWE-CI employs an Architect--Programmer dual-agent evaluation protocol: starting from the base commit, the agents execute a CI-loop that iteratively generates requirements, modifies source code, and runs tests, with the ultimate objective of passing all tests associated with the target commit. SWE-CI introduces \textbf{EvoScore} (\textbf{Evo}lution \textbf{Score}) as a proxy metric: it measures functional correctness on future modifications, so that agents whose earlier decisions facilitate subsequent evolution score higher, while those that accumulate technical debt see progressively declining performance.

We conduct extensive experiments with a total consumption of more than 10 billion tokens. Results reveal that, despite substantial progress on functional correctness, state-of-the-art models still struggle when tasked with sustaining code quality over extended evolution. We further provide comprehensive, fine-grained analyses of the evaluation results, offering valuable insights into the coding capabilities of LLM-based agents and demonstrating the distinctive diagnostic value of SWE-CI.

\section{Measuring the Agent's Ability to Maintain Codebase}

\subsection{Task formalization}
\label{sec:2.1}

\begin{figure}[t!] 
  \centering
  \includegraphics[width=0.9\textwidth, trim=0 200 140 0,clip]{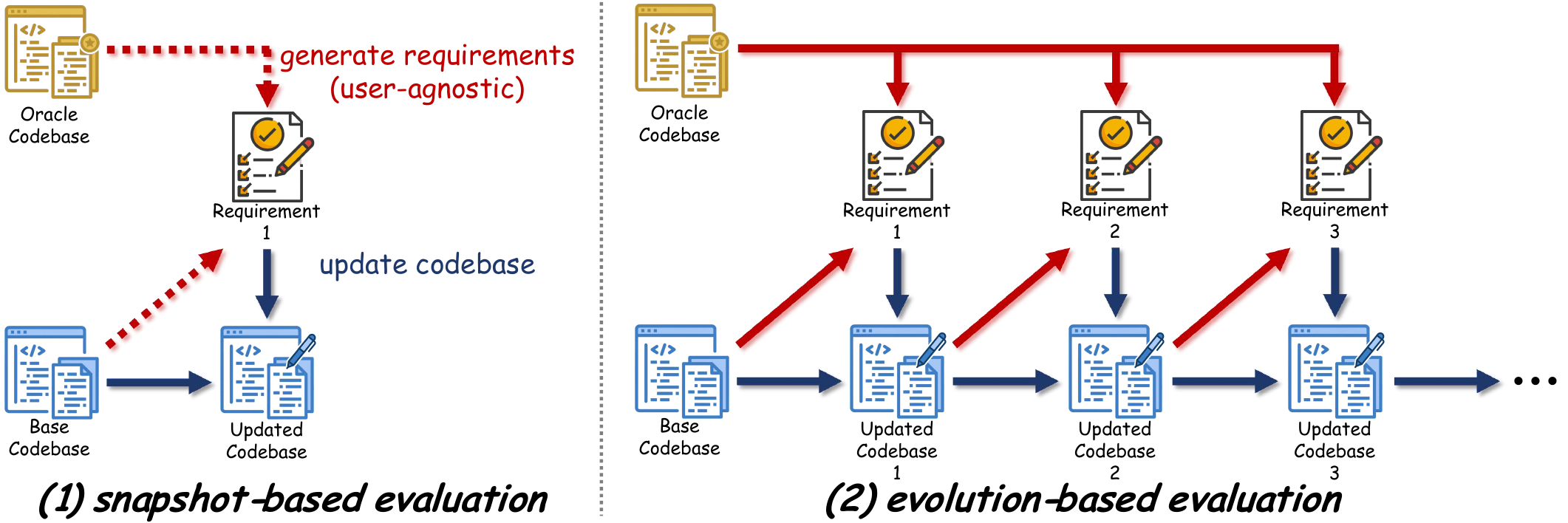} 
  \caption{Unlike previous benchmarks, SWE-CI proposes an evolution-based evaluation. The red and blue arrows represent the actions of functions $\mathsf{require}$ and $\mathsf{code}$, respectively. Dashed lines indicate processes that are unknown to the user.}
  \label{figure:1}
\end{figure}

We first establish a unified formalization for agentic coding tasks. Let $t$ denote a single unit test and $\mathcal{T} = \{t_1, t_2, \ldots, t_{|\mathcal{T}|}\}$ the set of all tests we are interested in. Let $\mathcal{C}$ denote the space of codebases and $\mathcal{R}$ the space of requirements. We further define two functions:
\begin{itemize}
    \item $\mathsf{require}_{\mathcal{T}}: \mathcal{C} \times \mathcal{C} \to \mathcal{R}$, which identifies the functional gap between two codebases with respect to $\mathcal{T}$, and produces a requirement document accordingly;
    \item $\mathsf{code}_{\mathcal{T}}: \mathcal{R} \times \mathcal{C} \to \mathcal{C}$, which modifies a codebase according to a given requirement and returns an updated codebase.
\end{itemize}
With these definitions, we observe that many prevailing coding benchmarks \cite{1,2,3,4,5,6} follow the \textit{snapshot-based} evaluation paradigm illustrated in Figure~\ref{figure:1}. In this paradigm, the requirement depends only on the base codebase $c_0$ and the ``golden'' codebase $c_\ast$, i.e., $r\equiv \mathsf{require}_{\mathcal{T}}(c_0, c_\ast)$. However, in this paper, we turn to consider an \textit{evolution-based} evaluation paradigm. The requirement in this paradigm is derived from the current codebase dynamically: $r_i = \mathsf{require}_{\mathcal{T}}(c_i, c_*)$, and the codebase is updated accordingly: $c_{i+1} = \mathsf{code}_{\mathcal{T}}(c_i, r_i)$. This iterative loop ensures that the consequences of earlier modifications propagate into subsequent iterations, making the agent's long-term decision quality observable.

\subsection{Normalized Change}

Most benchmarks treat passing all test cases as the gold standard for functional correctness. However, in software engineering, some features require long-term planning and incremental development rather than a one-shot implementation. Moreover, during evolution it is common for previously passing tests to be inadvertently broken --- a phenomenon known as \textit{regression}. We therefore need a finer-grained metric that reflects the current state of a codebase $c$, rather than a binary pass/fail verdict. To this end, we introduce the \textbf{normalized change}.

Let $n(c)$ be the number of test cases that $c$ passes:
\begin{equation}
n(c) = \sum_{t \in \mathcal{T}} \mathbb{I}(t, c),
\end{equation}
where the indicator function $\mathbb{I}(t, c)$ equals 1 if and only if unit test $t$ passes on codebase $c$, and 0 otherwise. The normalized change is then defined as:
\begin{equation}
a(c) = \left\{
\begin{aligned}
\frac{n(c) - n(c_0)}{n(c_\ast) - n(c_0)} & \,\,\,\,\text{if} \,\, n(c) \geq n(c_0), \\
\frac{n(c) - n(c_0)}{n(c_0)} & \,\,\,\,\text{else}.
\end{aligned}
\right.
\end{equation}

When the agent improves upon the base codebase ($n(c) \geq n(c_0)$), the numerator is normalized by the total gap $n(c_\ast) - n(c_0)$, so that $a(c) = 1$ if and only if the agent has closed the gap entirely. When the agent regresses below the baseline ($n(c) < n(c_0)$), the numerator is instead normalized by $n(c_0)$, so that $a(c) = -1$ corresponds to breaking every initially passing test. This asymmetric normalization is deliberate: regardless of how large or small $n(c_0)$ and $n(c_\ast)$ are, improvements and regressions are always expressed on a unified, comparable scale --- i.e., $a(c)\in[-1,1]$.

\subsection{EvoScore}

The ISO/IEC 25010 standard defines maintainability 
as the degree to which software can be modified 
effectively without introducing defects or degrading 
existing quality — simply put, more maintainable 
code is less likely to break future functionality. 
This yields a simple insight: maintainability can be 
revealed by tracking how functional correctness 
changes over time. Guided by this principle, given the sequence of codebases $(c_1, \ldots, c_N)$ obtained from $N$ iterations, we aggregate them into a single scalar, \textbf{EvoScore}, via a \textbf{future-weighted mean}:
\begin{equation}
e = \frac{\sum_{i=1}^N \gamma^{i}\,a(c_i)}{\sum_{i=1}^N \gamma^{i}}
\end{equation}
In EvoScore, we set $\gamma \geq 1$ so that later iterations receive \emph{greater} weight. The rationale directly mirrors the ISO definition: a truly maintainable codebase is one that remains easy to modify as evolution progresses. An agent that sacrifices short-term speed for a cleaner, more extensible design will be rewarded over one that rushes to pass early tests but accumulates technical debt that cripples subsequent evolution. When $\gamma = 1$, EvoScore reduces to the \textit{ average normalized change}; as $\gamma$ increases, the metric progressively favors long-term stability over immediate gains.

\section{SWE-CI}

\begin{figure}[t!]
  \centering
  \includegraphics[width=1\textwidth, trim=0 160 180 0,clip]{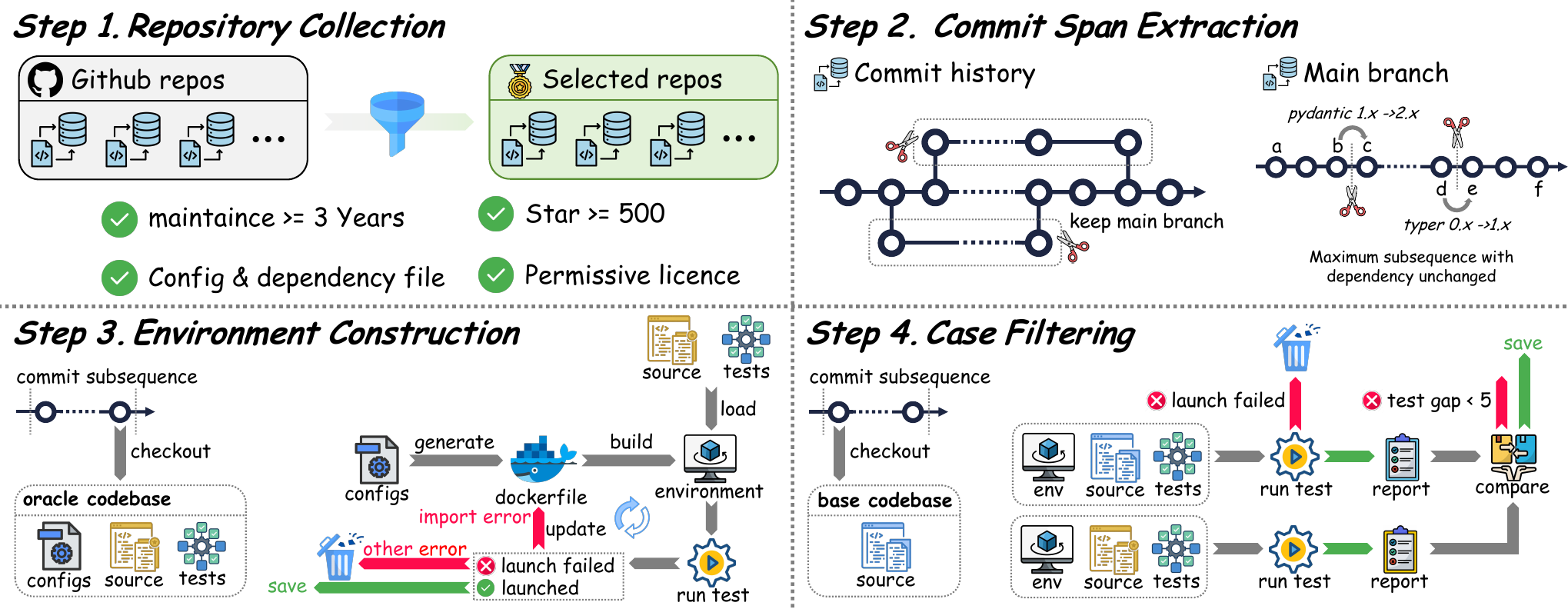} 
  \caption{Data curation process of SWE-CI.}
  \label{figure:2}
\end{figure}

\subsection{Data curation}
As shown in Figure~\ref{figure:1}, our goal is to obtain a number of base-codebase/oracle-codebase pairs and to let the agent iteratively evolve the former toward the latter, measuring its ability to maintain code throughout this process. Each such pair can be viewed as two chronologically ordered commits within the same repository. Concretely, the construction of SWE-CI is carefully orchestrated as follows:

\paragraph{Step 1: Repository Collection.}
Unlike SWE-Bench and similar benchmarks that draw exclusively from a handful of well-known open-source prjects, we cast a wider net by searching across all Python repositories on GitHub. We then apply the following filtering criteria: (1) the repository has been actively maintained for at least three years; (2) it has accumulated more than 500 stars; (3) it contains configuration and dependency files (e.g., \texttt{pyproject.toml} and lockfiles) as well as a suite of unit tests; and (4) it is released under a permissive license such as MIT or Apache-2.0. After applying these filters, 4,923 repositories remain.

\paragraph{Step 2: Commit Span Extraction.}
For each surviving repository, we retain only its main branch, reducing the history to a linear sequence of commits. We then compare the dependencies of consecutive commits along this sequence and identify all maximal subsequences within which the dependencies remain unchanged. The two endpoints of every such subsequence naturally form a candidate base/oracle pair. We further discard pairs whose total number of modified lines of code is below 1,000, as such pairs represent insufficient evolutionary distance. This process yields 8,311 candidate pairs.

\paragraph{Step 3: Environment Construction.}
For each candidate pair, we automatically generate a Dockerfile based on the configuration and dependencies of the oracle codebase and snapshot the resulting runtime environment. We then execute the oracle codebase's unit test suite within this environment to verify its correctness. To improve data retention, we introduce a self-repair mechanism: whenever the test suite fails to launch due to a missing dependency, we detect the failure and dynamically inject the required dependency into the Dockerfile to build a new environment. This mechanism substantially increases the number of viable candidate pairs. Pairs whose failures stem from other reasons are discarded. After this step, 1,458 candidate pairs and their runtime environment snapshots remain.

\paragraph{Step 4: Case Filtering.}
Finally, we apply three further rounds of filtering to ensure the quality of the final dataset. First, within the runtime environment snapshot constructed in Step 3, we run the oracle codebase's test suite against the base codebase. Any candidate whose tests fail to launch is discarded. Second, we compare the test reports produced by the base and oracle codebases on the same test suite; candidates for which the difference in the number of passing tests is fewer than five are removed. After these two automated filters, 137 candidates remain. In the last round, we rank the surviving candidates by their time span and number of intervening commits, and select the top 100 to form the final SWE-CI benchmark.

The final SWE-CI benchmark comprises 100 samples drawn from 68 distinct repositories. On average, each base/oracle pair is separated by 233 days and 71 consecutive commits of real-world development history. In every pair, the transition from the base to the oracle codebase involves at least 500 lines of modified source code, excluding changes to test files. Each sample is shipped with the complete source code and a pre-built Docker environment to ensure reproducibility. These statistics confirm that SWE-CI captures substantial, long-term evolutionary episodes rather than trivial incremental changes.

\subsection{Dual-agent evaluation protocol}
\begin{figure}[t!]
  \centering
\includegraphics[width=1\textwidth, trim=0 240 530 0,clip]{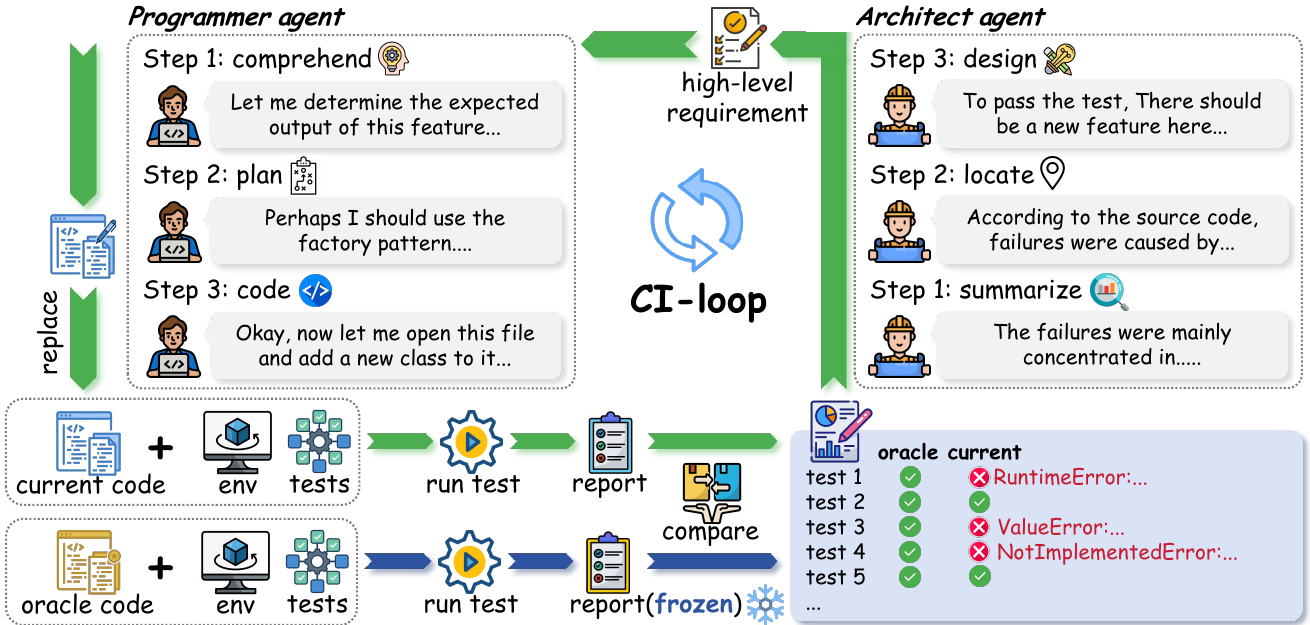} 
  \caption{SWE-CI uses an architect-programmer dual-agent workflow to model the continuous integration cycle of professional software teams in the real world.}
  \label{figure:3}
\end{figure}
As described in Figure \ref{figure:1}, SWE-CI adopts evolution-based evaluation. To support this setting, we introduce an Architect-Programmer dual-agent protocol. The Architect identifies functional gaps and issues requirements; the Programmer implements them. Their collaboration reproduces the CI loop in real-world development, enabling fine-grained observation of how well agents maintain code.

\paragraph{Architect agent.} 

Based on the test gap between the current code and the oracle code, the Architect is tasked with producing a high-level requirements document in natural language. We prompt the architect to organize its behavior into three steps: \ding{182} \textbf{Summarize}. Architect reviews all failing tests, identifies root causes, and identifies source code files that need further inspection; \ding{183} \textbf{Locate}. Architect examines the source code and attributes failures to concrete deficiencies in the current implementation; \ding{184} \textbf{Design}. Based on these deficiencies, architect devises an improvement plan and produces the final requirements document. Two writing conventions are further imposed on requirement document. \ding{192} \textit{Incremental}. The document should contain no more than five of the most urgent requirements, avoiding the pitfall of over-designing in a single iteration. \ding{193} \textit{High-level}. the requirements should focus on describing expected behavior of code using natural language, leaving concrete implementation choices to the programmer. The core purpose of these specifications is to ensure that requirement documents meet the needs of real-world continuous integration processes.

\paragraph{Programmer.} The programmer's responsibility is to maintain the code according to the requirements document. Programmer behavior is also standardized into three steps: \ding{182} \textbf{Comprehend}. Programmers understand high-level language requirements and translate them into explicit code specifications. \ding{183} \textbf{Plan}. Programmers plan the programming effort required to implement these specifications. \ding{184} \textbf{Code}. Programmers put these plans into practice and try to fulfill the requirements.

In this protocol, the Programmer is driven by the requirements document rather than directly by the test gap --- a deliberate design choice that aligns with the rapid iteration philosophy of continuous integration. To this end, the Architect is required to distill the most pressing requirements from the full set of failures, allowing the Programmer to focus on fast, targeted development without being overwhelmed by the full scope of changes.

\section{Experiments}

\subsection{Experiment setting}
We use \verb|pytest| and \verb|pytest-json-report| as the testing framework, with a timeout of 3600 seconds per test run. iFlow CLI \cite{10} serves as the default harness, and the maximum number of iterations in the dual-agent evaluation protocol is set to 20. Unless otherwise specified, the Architect Agent and the Programmer Agent share the same underlying base model.

\subsection{Maintainability}

\begin{figure}[h!]
  \centering
\includegraphics[width=1\textwidth, trim=0 0 0 0,clip]{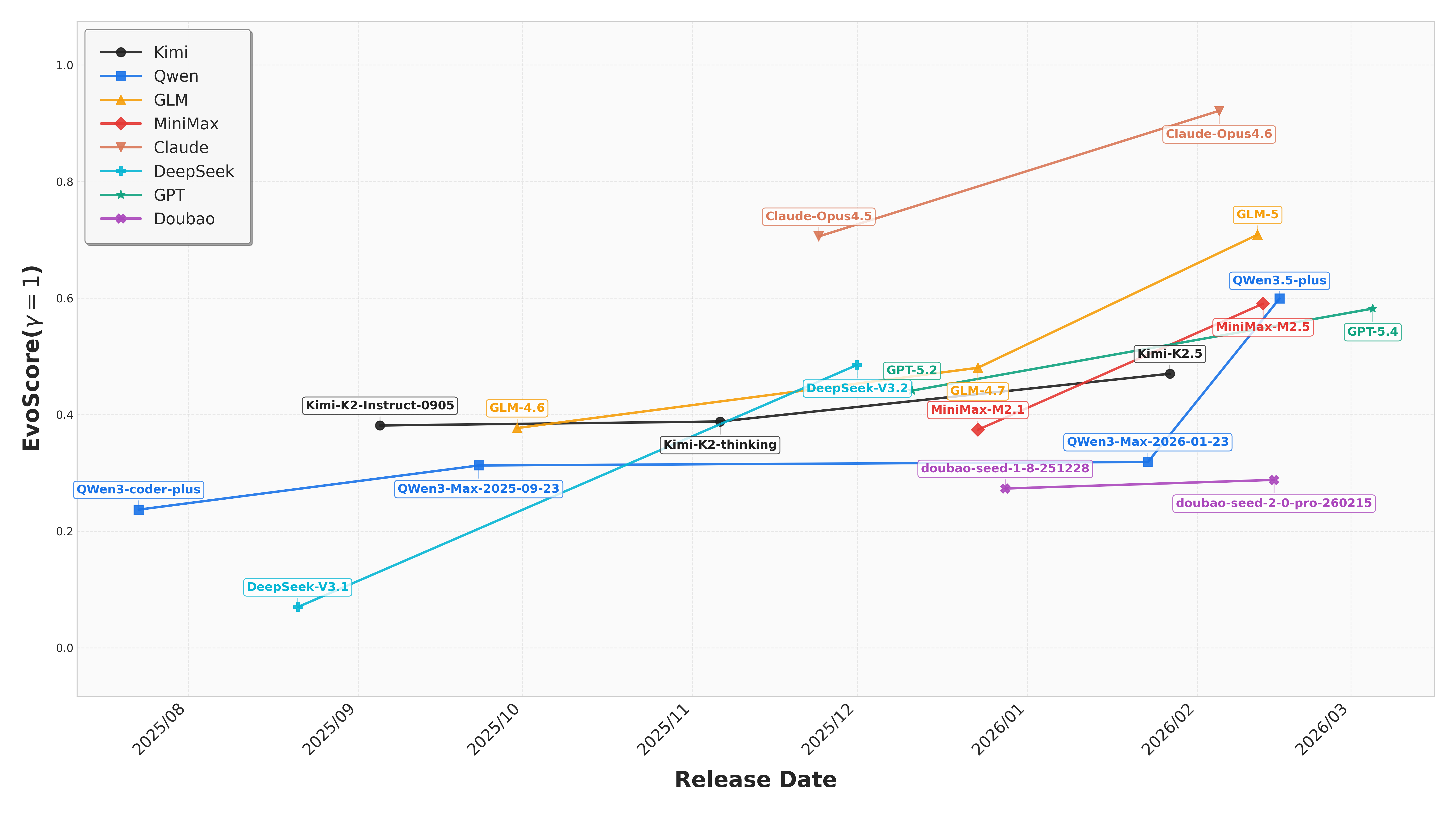} 
  \caption{The EvoScore variation of state-of-the-art models from 8 providers on SWE-CI.}
  \label{figure:4}
\end{figure}

\paragraph{Observation 1: The code maintenance capabilities of LLMs are advancing at an accelerating pace (Figure \ref{figure:4}).} Extensive evaluation of 20 models from 8 providers reveals a consistent pattern: within the same provider, newer models always achieve higher scores, with models released after 2026 showing markedly larger gains than their predecessors. This suggests that the code capabilities of current LLMs are rapidly evolving beyond static bug-fixing toward sustained, long-term code maintenance. Among all evaluated models, the Claude Opus series demonstrates a commanding lead throughout the entire observation period, with GLM-5 also standing out as a strong performer.

\begin{figure}[h!]
  \centering
\includegraphics[width=1\textwidth, trim=0 0 0 0,clip]{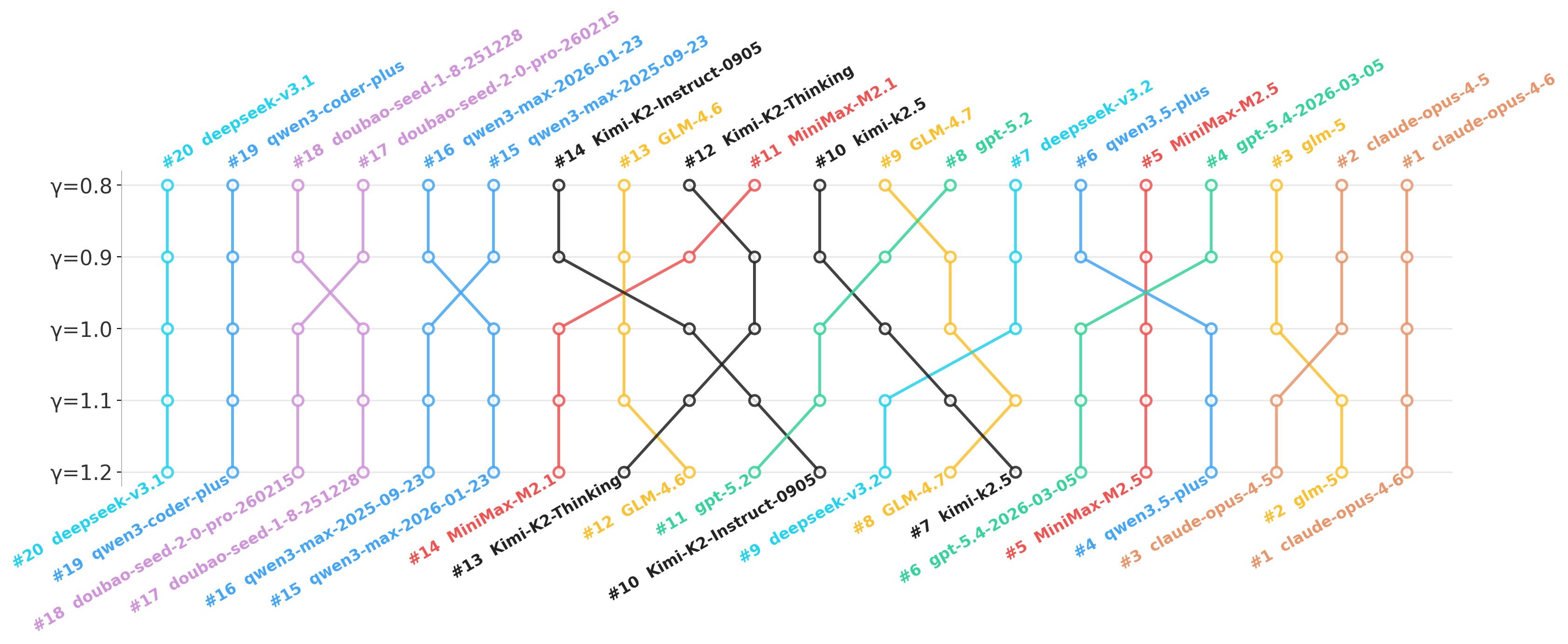} 
  \caption{The model's EvoScore ranking changes as $\gamma$ increases. When $\gamma > 1$, higher-ranking models indicate better codebase maintenance.}
  \label{figure:5}
\end{figure}

\paragraph{Observation 2: Different providers place varying degrees of emphasis on code maintainability (Figure \ref{figure:5}).} We vary the value of $\gamma$ to examine how model rankings shift accordingly. When $\gamma < 1$, EvoScore assigns higher weights to earlier iterations, favoring models that prioritize immediate gains from code modification. Conversely, when $\gamma > 1$, later iterations are rewarded, giving an advantage to models that optimize for long-term improvement, i.e., prioritize code maintainability. We find that preferences vary considerably across providers, while models within the same provider tend to exhibit consistent tendencies. Specifically, MiniMax, GPT, and DeepSeek show a preference for short-term gains, with their rankings declining notably as $\gamma$ increases, whereas Kimi, GLM, and Qwen lean toward long-term maintainability, with their rankings rising accordingly. Doubao and Claude, by contrast, remain relatively stable across different $\gamma$ settings, as their overall performance advantage or disadvantage is sufficiently pronounced to absorb the effect of reweighting. We conjecture that these divergences reflect differences in training strategies adopted by different providers, while the relative consistency within each provider suggests that their internal training pipelines remain largely stable.

\subsection{Regression}

\begin{figure}[h!]
  \centering
\includegraphics[width=1\textwidth, trim=0 0 0 0,clip]{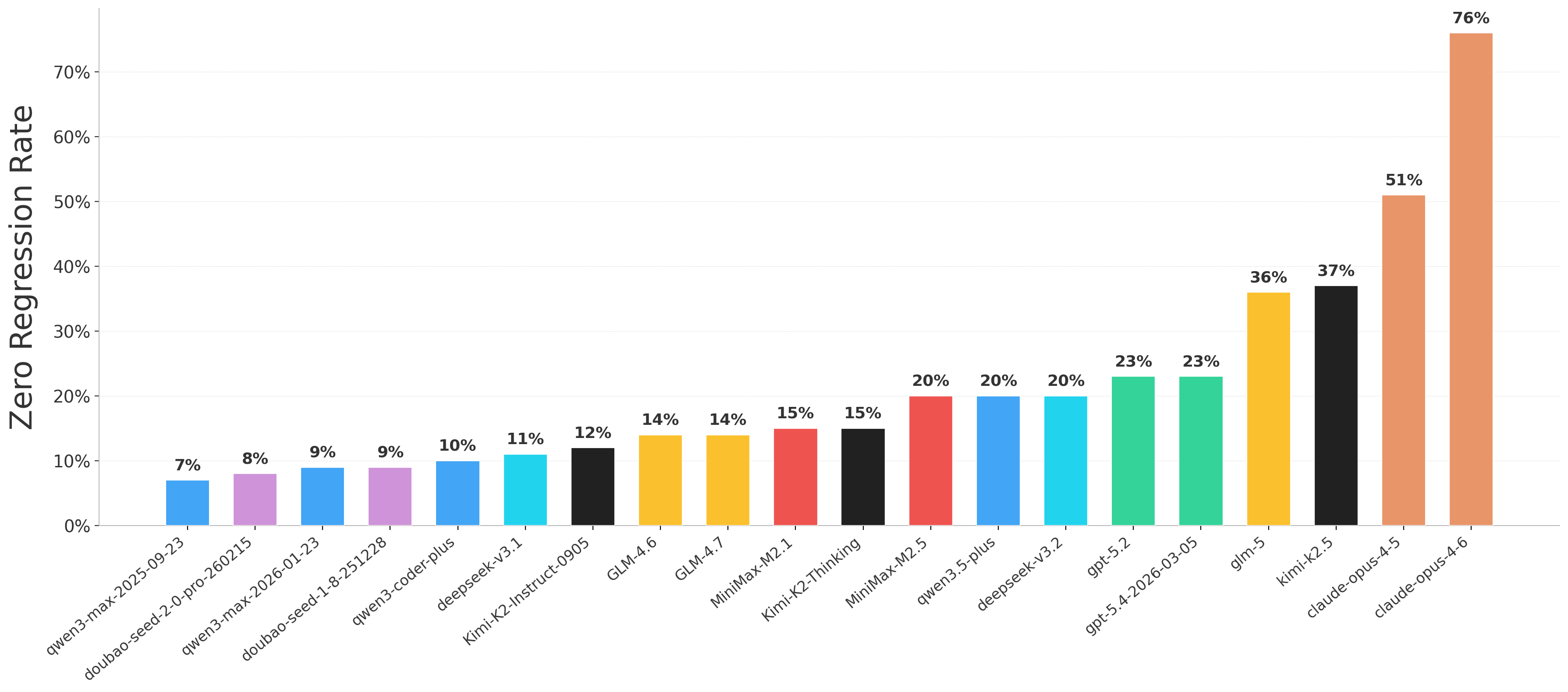} 
  \caption{All models are sorted from smallest to largest by the zero regression rate.}
  \label{figure:6}
\end{figure}

\paragraph{Observation 3: Current LLMs still fall short in controlling regressions during long-term code maintenance (Figure \ref{figure:6}).} A regression occurs when a code change causes a previously passing test to fail. In long-term maintenance, regressions are particularly damaging as their effects compound across successive iterations. We measure the zero-regression rate, defined as the proportion of samples in which no regression occurs throughout the entire maintenance process, as an indicator of model stability in continuous maintenance scenarios. Experimental results show that most models achieve a zero-regression rate below 0.25, with only the two Claude-opus models exceeding 0.5. This indicates that current LLMs still struggle to reliably avoid regressions across long-term, multi-round code maintenance, despite their strong performance on snapshot-based tasks.

\begin{figure}[h!]
  \centering
  \begin{subfigure}{0.48\textwidth}
    \centering
    \includegraphics[width=\textwidth]{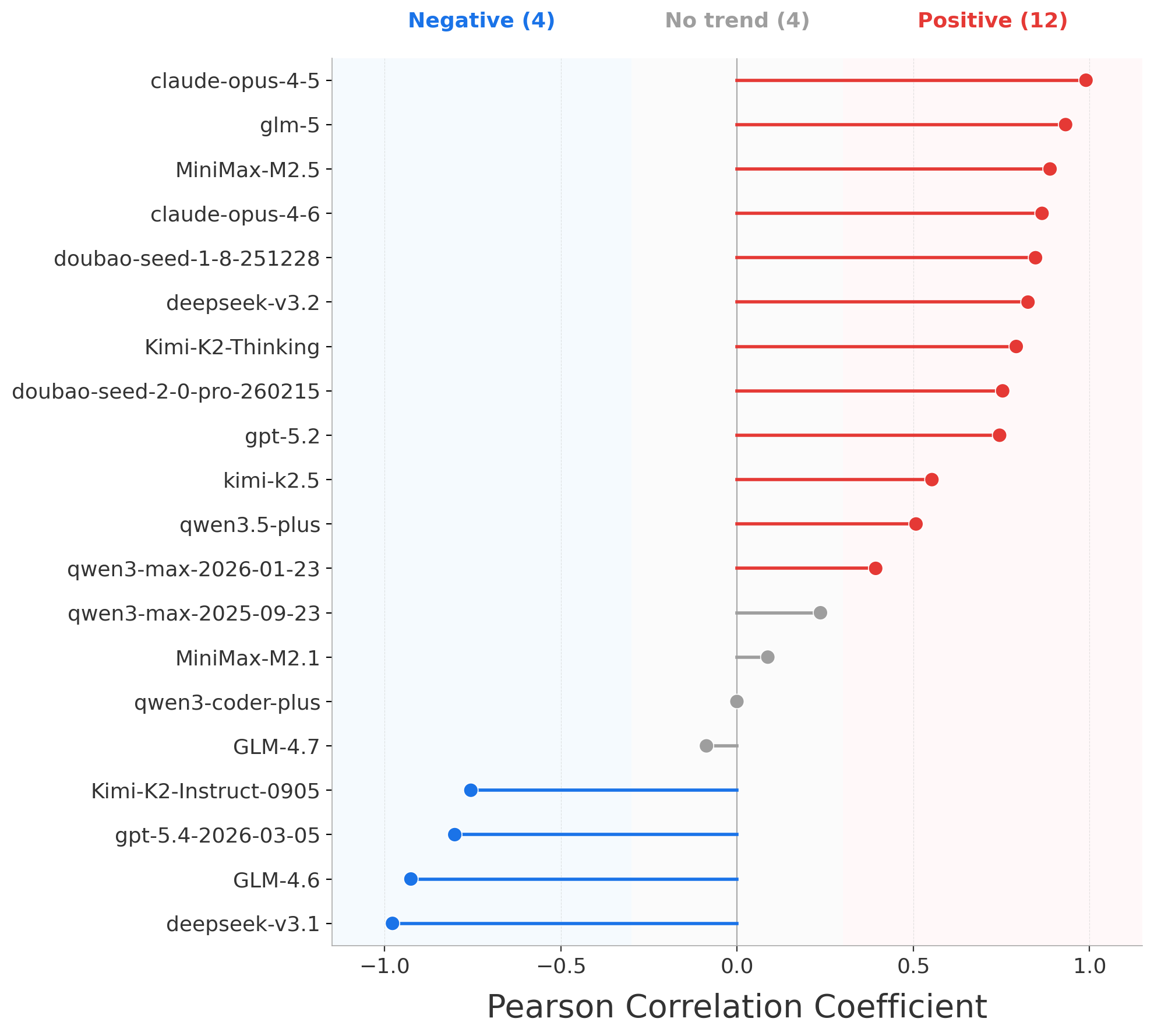}
  \end{subfigure}
  \hfill
  \begin{subfigure}{0.48\textwidth}
    \centering
    \includegraphics[width=\textwidth]{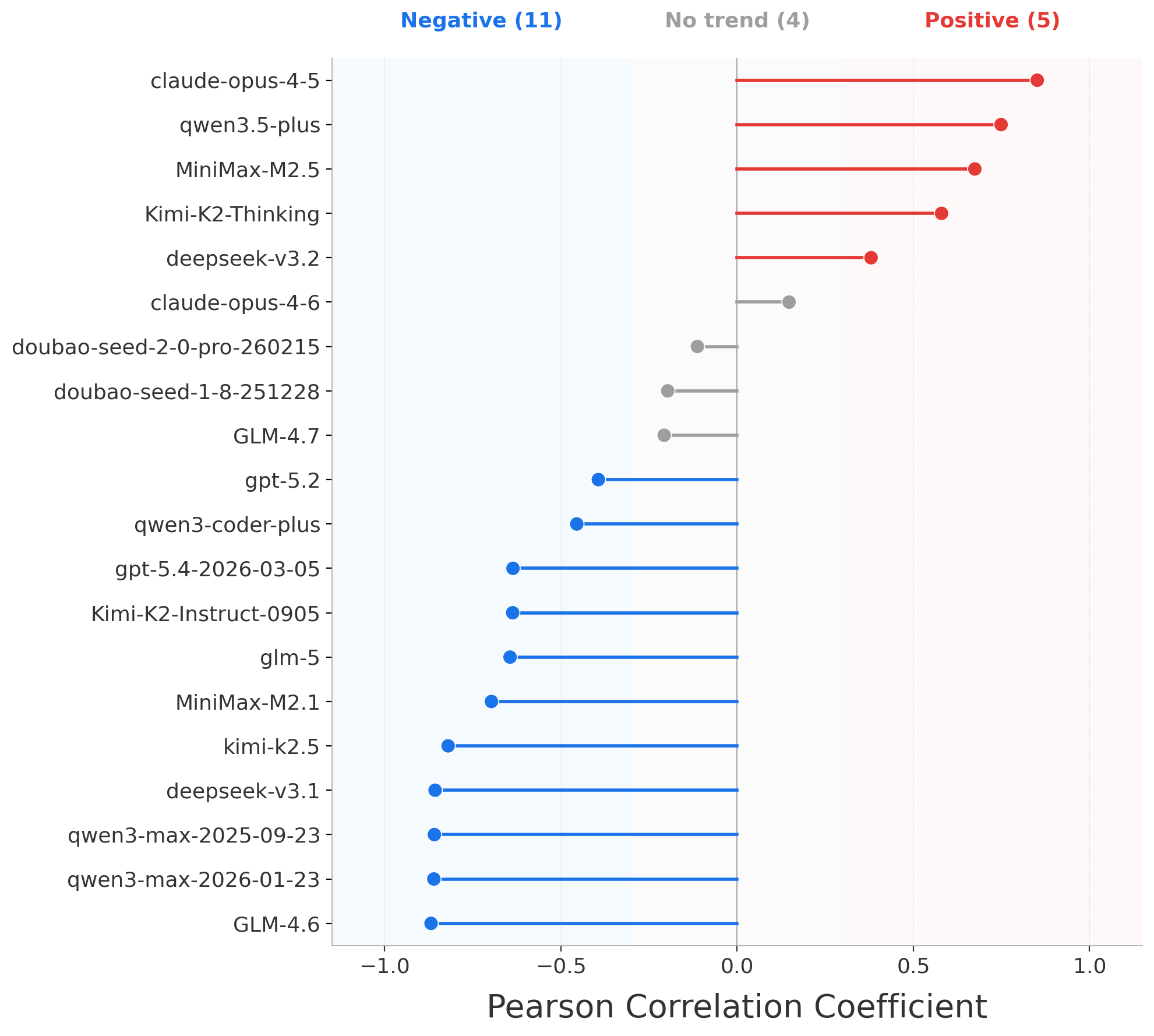}
  \end{subfigure}
    \caption{Pearson correlation of regression rate (left) and magnitude (right) with number of iterations.}
    \label{figure:7}
\end{figure}

\paragraph{Observation 4: As iterations progress, regression rate increases for most models while regression magnitude decreases (Figure \ref{figure:7}).} We further analyze how regression behavior evolves across iterations by computing the Pearson correlation between the number of iterations and two regression metrics: regression rate and regression magnitude. Results show that the two metrics exhibit divergent trends across most models. For regression rate (left), 12 out of 20 models show a positive correlation with iteration count, indicating that regressions become increasingly frequent as maintenance progresses. For regression magnitude (right), however, 11 out of 20 models show a negative correlation, suggesting that each individual regression affects a shrinking portion of the test suite over time. We conjecture that this divergence reflects the natural structure of long-term maintenance tasks: as easy requirements are resolved in early iterations, models increasingly resort to trial-and-error on harder problems in later iterations, resulting in more frequent but more localized modifications that raise regression rate while reducing regression magnitude.

\subsection{Coding style}

\begin{figure}[h!]
  \centering
  \begin{subfigure}{0.48\textwidth}
    \centering
    \includegraphics[width=\textwidth, trim=0 0 0 0, clip]{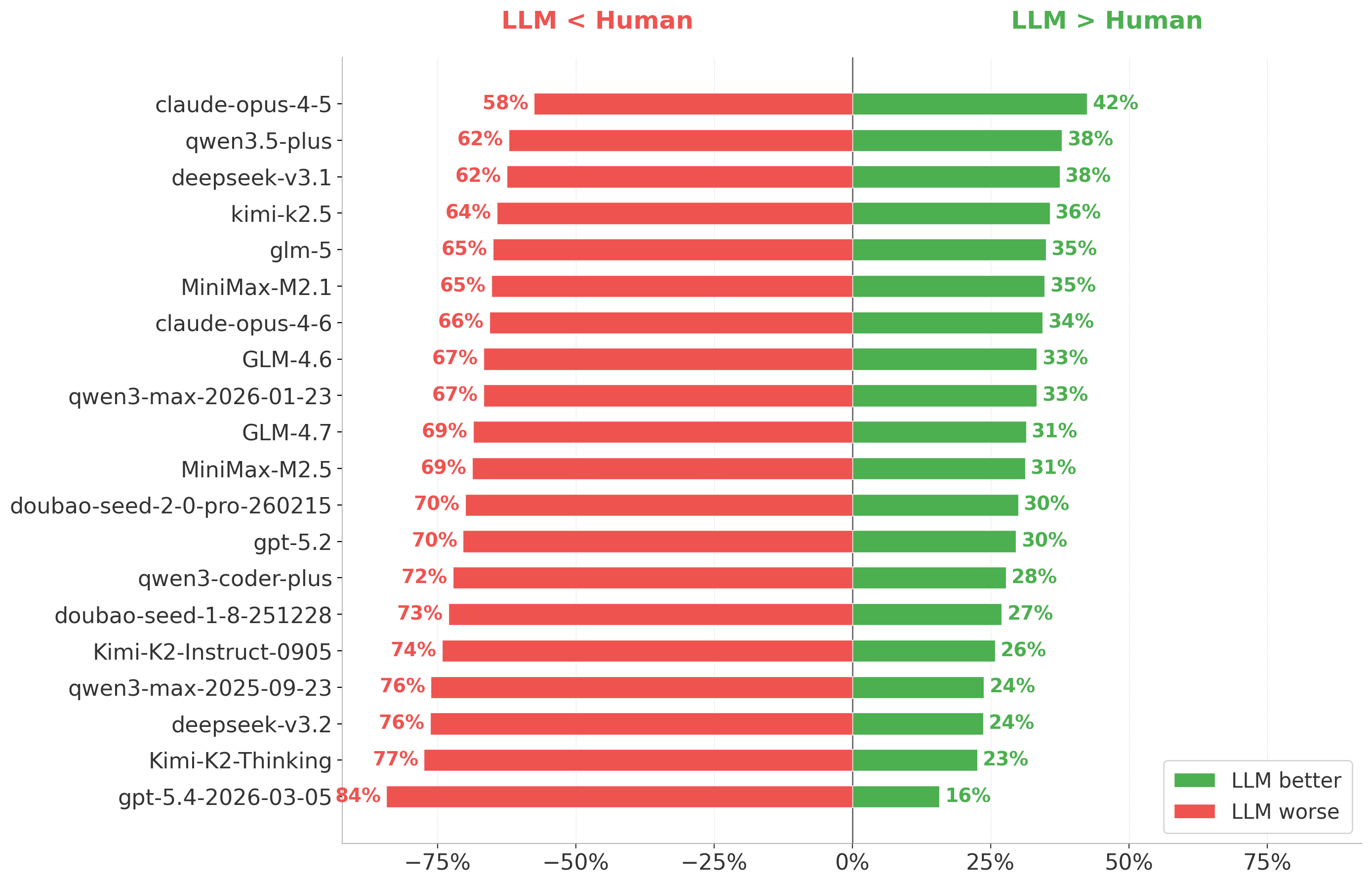}
  \end{subfigure}
  \hfill
  \begin{subfigure}{0.48\textwidth}
    \centering
    \includegraphics[width=\textwidth, trim=0 0 0 0, clip]{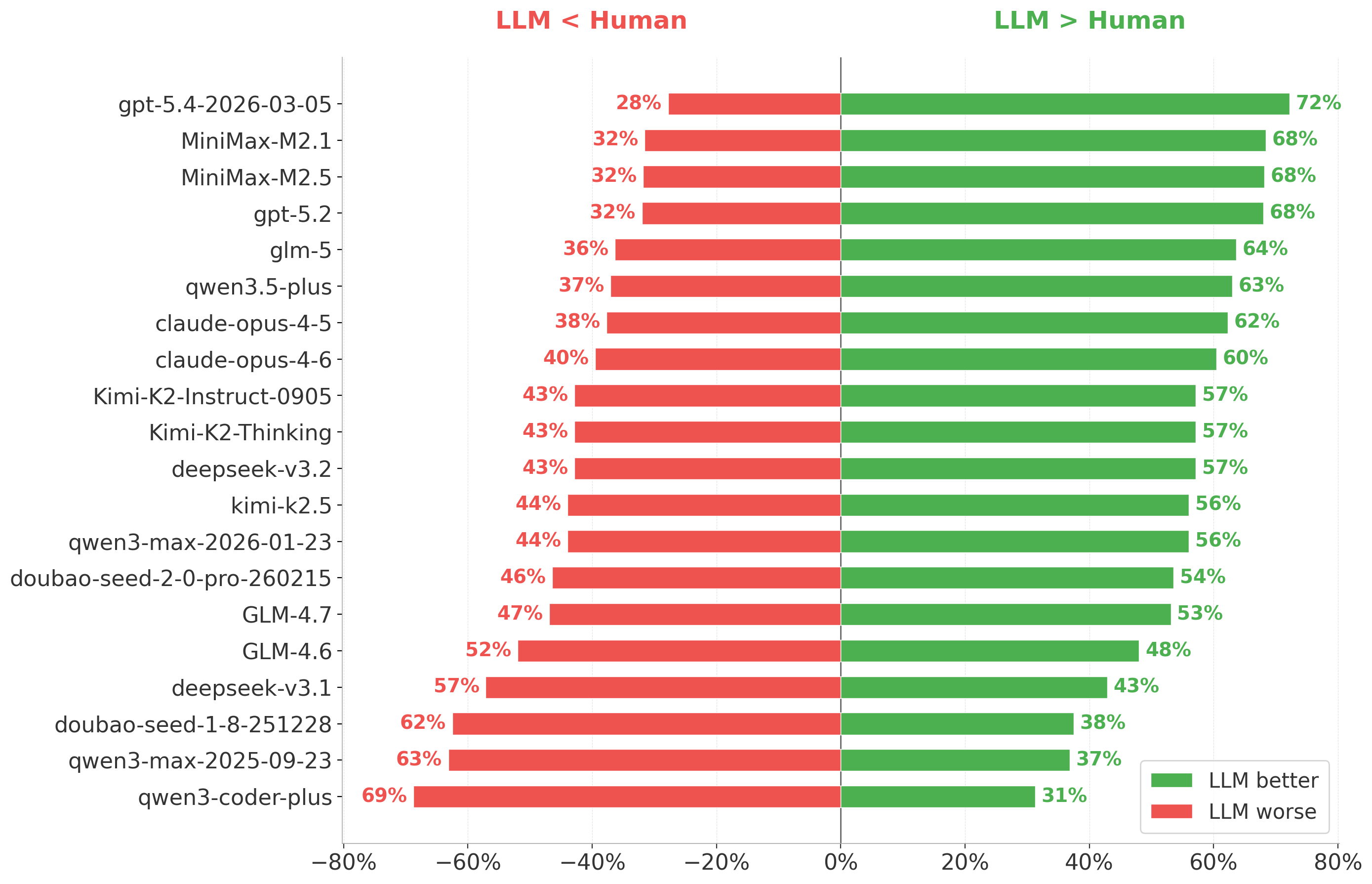}
  \end{subfigure}
  \caption{Win/loss rate of LLM-generated code against human oracle code in terms of MI score (left) and Pylint score (right) on successfully solved problems.}
  \label{figure:8}
\end{figure}

\paragraph{Observation 5: LLMs excel at surface-level code style but fall short in true code maintainability (Figure \ref{figure:8}).} We compare LLM-generated code against human oracle solutions on successfully solved problems, measuring the win/loss rate across two static code metrics: Pylint score and MI score. Specifically, Pylint score reflects adherence to surface-level coding conventions such as line length and naming rules, while MI score captures deeper maintainability through indicators such as cyclomatic complexity. Across all 20 evaluated LLMs, a striking pattern emerges: while the majority of LLMs (15 out of 20) \textit{outperform} human oracle code on Pylint score, all 20 LLMs \textit{underperform} on MI score. This divergence suggests that LLMs are more adept at learning explicit, local coding conventions than implicit, global code quality criteria, which may reflect an inherent limitation in how LLMs model code structure. In other words, LLM-generated code tends to be \textit{superficially clean but intrinsically complex}, whereas human-written oracle code, though less strictly formatted, achieves higher maintainability through simpler and more elegant logic.

\begin{figure}[h!]
  \centering
\includegraphics[width=1\textwidth, trim=0 0 0 0,clip]{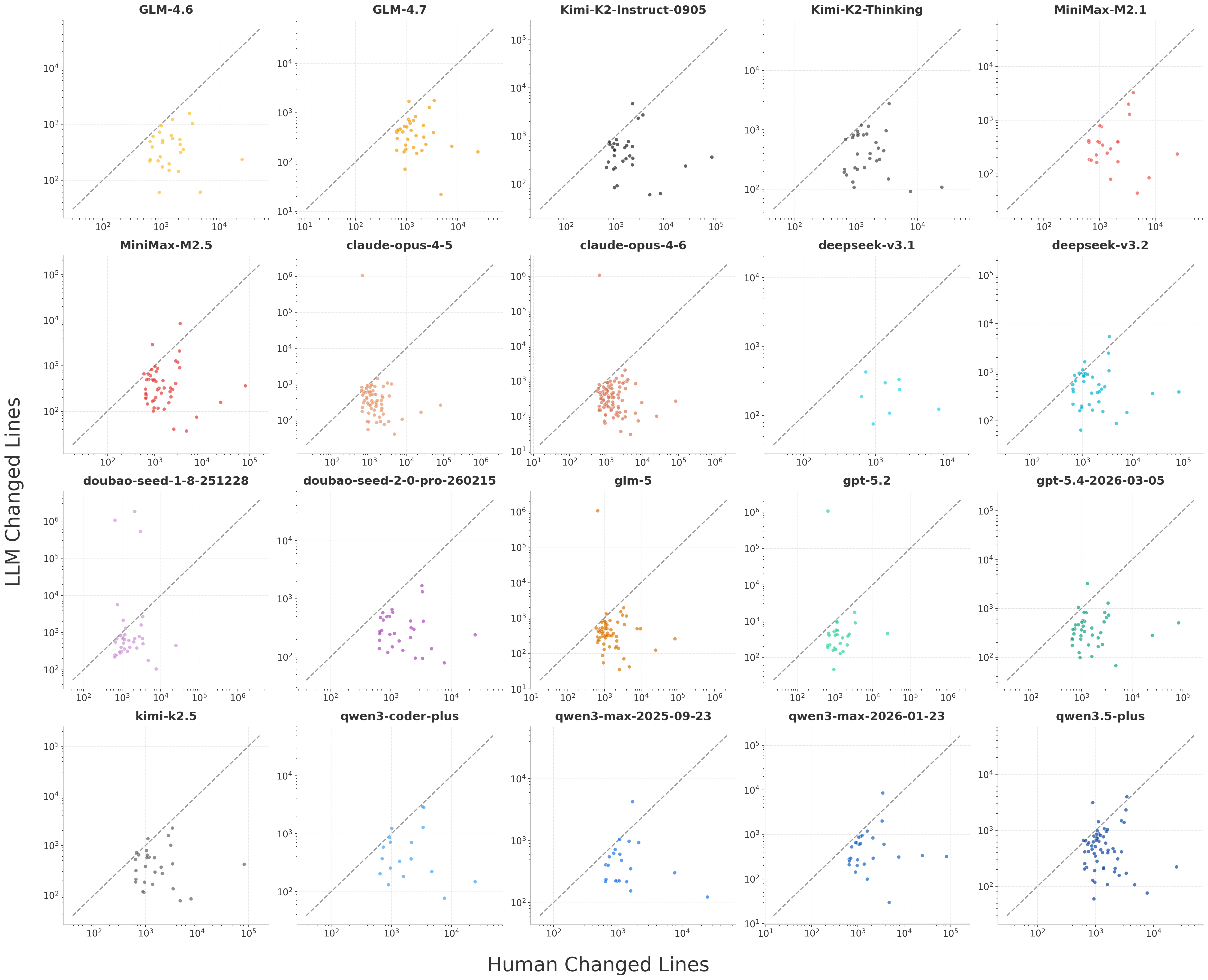} 
  \caption{Scatter plots comparing changed lines between LLM-generated and human oracle solutions across successfully solved problems.}
  \label{figure:9}
\end{figure}

\paragraph{Observation 6: LLMs generate more concise patches than humans, yet at the cost of maintainability (Figure \ref{figure:9}).} We further examine the number of changed lines between LLM-generated and human oracle solutions across all successfully solved problems. Across all 20 evaluated LLMs, points consistently cluster \textit{below} the diagonal, indicating that LLMs universally produce more concise patches than their human counterparts. This observation further reinforces Observation 5: when considered together, LLMs achieve lower MI scores with fewer changed lines, whereas human solutions, though more verbose, attain higher maintainability. This suggests that the additional lines written by humans are not redundant, but rather purposeful investments 
in code quality through practices such as abstraction, encapsulation, and modularization. LLMs, by contrast, appear to optimize for the minimal changes necessary to satisfy immediate task requirements, without investing in the broader code structure that characterizes maintainable software engineering practice.

\section{Conclusion}

We present SWE-CI, a repository-level benchmark that operationalizes maintainability as functional correctness on future modifications — making visible what snapshot-based benchmarks cannot: the cumulative consequences of an agent's design decisions as the codebase evolves. Extensive experiments across 20 models from 8 providers reveal that current LLMs still struggle to sustain code quality over extended evolution, particularly in controlling regressions. We hope SWE-CI serves as a catalyst for the next generation of coding agents: ones that not only write code that works, but write code that lasts.

\clearpage

\section*{Acknowledgements}

The authors thank colleagues from Alibaba Group, including Jianan Ye, Cecilia Wang, Yijie Hu, Zongwen Shen, and Mingze Li, for their valuable suggestions and feedback on this paper.

\bibliographystyle{nips}
\bibliography{references}
\clearpage

\appendix

\section{Related Work}
\paragraph{Function-Level Code Generation Benchmarks.} Early benchmarks for evaluating LLM code capabilities focused on the function-level synthesis paradigm. HumanEval \cite{1} introduced 164 Python programming problems paired with unit tests, establishing the widely adopted pass@k metric. MBPP \cite{2} extended this paradigm with 974 crowd-sourced entry-level tasks. APPS \cite{11} and CodeContests \cite{12} further raised the difficulty bar by drawing from competitive programming contests. EvalPlus \cite{13} augmented the test suites of HumanEval and MBPP, revealing that model performance had been systematically overestimated. CodeXGLUE \cite{14} and MultiPL-E \cite{15} broadened evaluation to multiple programming languages and tasks. DS-1000 \cite{16} targeted data science workflows involving real-world libraries such as NumPy and Pandas. BigCodeBench \cite{17} stressed practical coding ability by requiring models to compose function calls across 139 real-world libraries. LiveCodeBench \cite{3} addressed data contamination through a continuously updated protocol sourcing problems from live competitive programming platforms. Together, these benchmarks have established a rigorous foundation for function-level evaluation, yet their scope remains confined to isolated tasks --- falling short of capturing the complexity of real-world software engineering.

\paragraph{Repository-Level Software Engineering Benchmarks.} To move beyond function-level evaluation, a second generation of benchmarks situates agents within realistic software engineering contexts. RepoBench \cite{18} introduced repository-level code completion tasks requiring models to retrieve and leverage cross-file context. SWE-bench \cite{4} marked a significant step forward by requiring agents to resolve real GitHub issues within complete repository contexts, operationalizing the ``Issue-to-PR'' paradigm; its subsequent variants — SWE-bench Verified \cite{4}, SWE-bench Multimodal \cite{19}, and SWE-bench Pro \cite{20} — further improved evaluation reliability, extended the paradigm to visual artifacts, and scaled to enterprise-level complexity respectively. DevBench \cite{21} broadened the scope to the full software development lifecycle, from requirements analysis through implementation and testing. At the agent interaction level, AgentBench \cite{22} evaluated agents across multiple interactive environments, Terminal-bench \cite{5} targeted hard, realistic tasks in command-line interfaces, $\tau$-bench \cite{6} probed multi-turn tool use in real-world service domains, and $\tau^2$-bench \cite{23} extended this setting to dual-control environments where both agent and user actively interact with a shared world. Together, these benchmarks have substantially advanced the evaluation of agents in complex software engineering settings, yet they universally adopt a snapshot-based protocol in which the agent addresses a single, static requirement in one pass — leaving long-term code maintainability beyond their reach.

\paragraph{Long-Horizon and Evolution-Aware Benchmarks.} A more recent line of work attempts to move beyond single-issue, snapshot-based evaluation toward longer-horizon and more dynamic settings. InterCode \cite{25} introduced interactive code execution environments where agents iteratively refine solutions through execution feedback. Commit0 \cite{24} challenged agents to implement entire libraries from scratch given API specifications and multi-stage feedback. LongCLI-Bench \cite{26} targeted long-horizon CLI tasks across four engineering categories with regression-aware evaluation. SWE-Bench-CL \cite{27} organized GitHub issues into chronological sequences to evaluate continual learning and catastrophic forgetting in coding agents. SWE-EVO \cite{28} constructed evolution tasks from real release histories, requiring agents to implement multi-file modifications while preserving existing functionality. Despite these advances, none explicitly model the cumulative degradation of code quality across iterative development cycles — the central concern our benchmark is designed to address.

\paragraph{Software Maintainability and Evolution.} The challenge of maintaining and evolving software systems has long been recognized in software engineering research. Lehman's Laws \cite{7} established that software complexity inevitably grows over time, while Cunningham's technical debt metaphor \cite{30} formalized how short-term shortcuts accumulate into long-term productivity losses. Empirically, software maintenance has been shown to account for the majority of total software lifecycle costs \cite{29}. Yet despite this long-standing recognition, existing benchmarks evaluate agents on whether code works — not on whether it remains maintainable as the codebase evolves. Our benchmark is designed to close this gap, providing the first evaluation framework that explicitly measures the maintainability of agent-generated code under realistic, iterative development conditions.

\section{Additional Experiments}

\subsection{Solved rate}

\begin{figure}[h!]
  \centering
\includegraphics[width=1\textwidth, trim=0 0 0 0,clip]{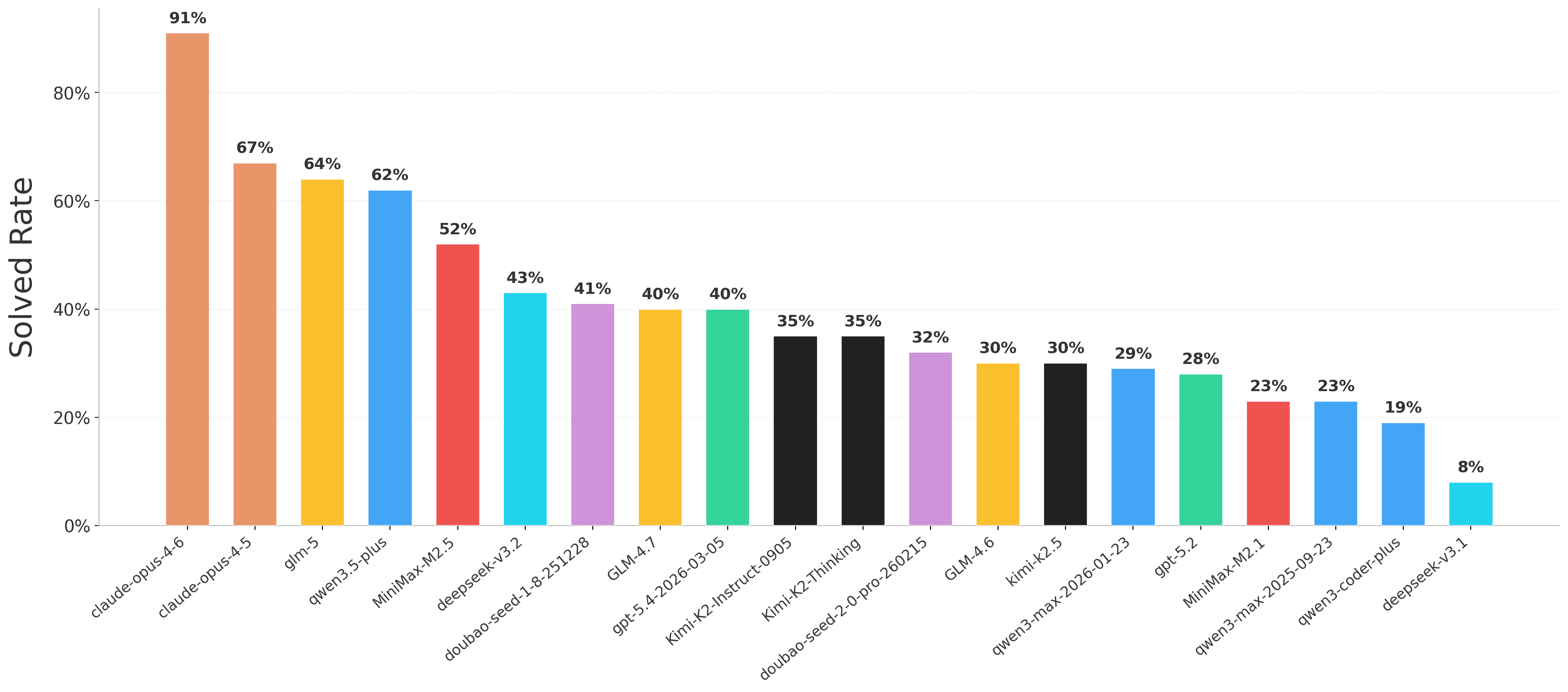} 
  \caption{The solved rate of all evaluated models, sorted in descending order.}
  \label{figure:10}
\end{figure}

\paragraph{Observation 7: Most models resolve fewer than half of all requirements, consistent with EvoScore rankings (Figure \ref{figure:10}).} The solved rate measures the proportion of samples in which all requirements are successfully fulfilled by the end of the maintenance process, and can be interpreted as a limiting case of EvoScore when $\gamma \rightarrow \infty$. The ranking of models largely aligns with their EvoScore rankings in Figure \ref{figure:4}, suggesting that EvoScore effectively captures long-term task completion ability. Nevertheless, most models achieve a solved rate below 50\%, further confirming the difficulty of SWE-CI as a long-term code maintenance benchmark.

\subsection{Average turns}

\begin{figure}[h!]
  \centering
\includegraphics[width=1\textwidth, trim=0 0 0 0,clip]{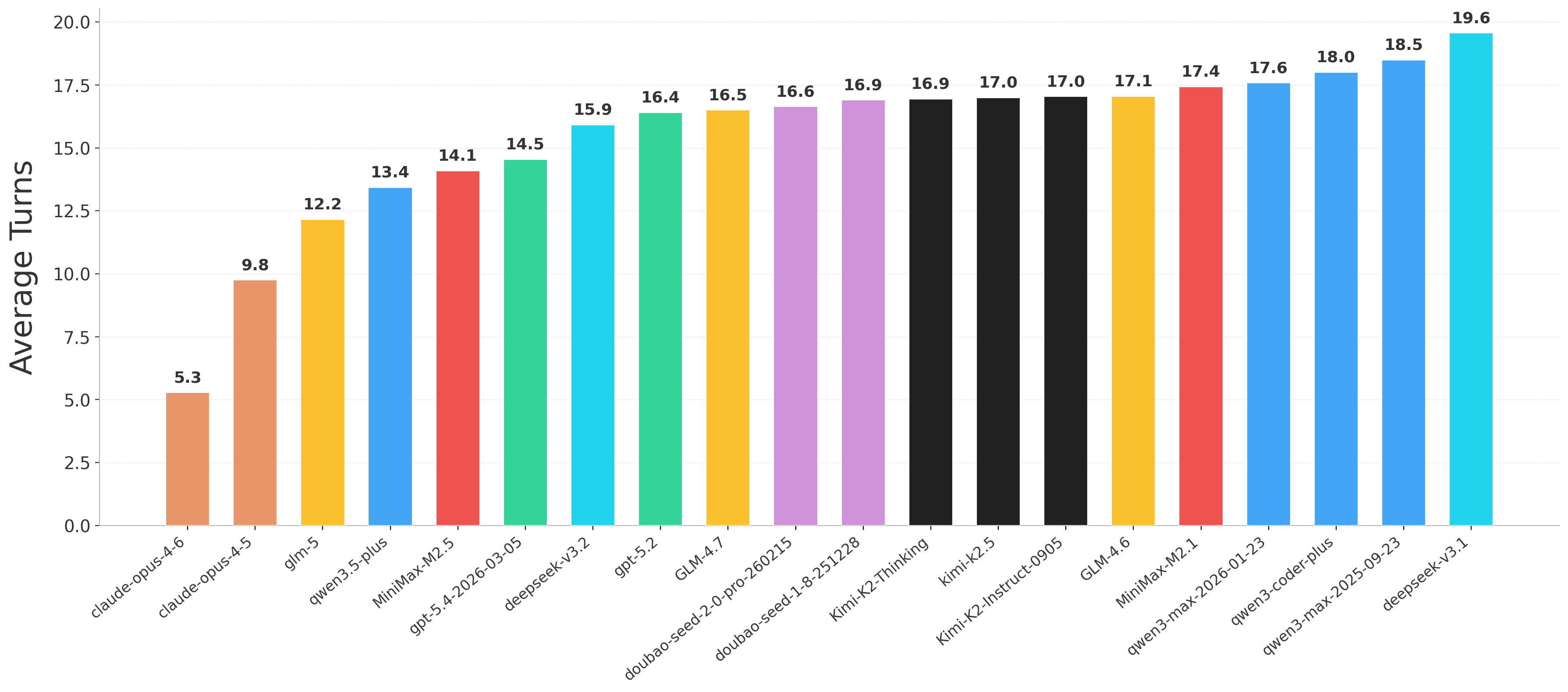} 
  \caption{The average number of turns consumed per task across all evaluated models, sorted in ascending order.}
  \label{figure:11}
\end{figure}

\paragraph{Observation 8: Additional turns do not compensate for limited model capability (Figure \ref{figure:11}).} We report the average number of turns consumed per task across all 20 models, with a maximum of 20 turns allowed. Weaker models tend to exhaust the turn budget without resolving the remaining requirements, whereas stronger models terminate early having completed the tasks efficiently. This suggests that in long-term code maintenance, the bottleneck lies in model capability rather than the number of attempts.

\subsection{Maximum relative changes}

\begin{figure}[h!]
  \centering
\includegraphics[width=1\textwidth, trim=0 0 0 0,clip]{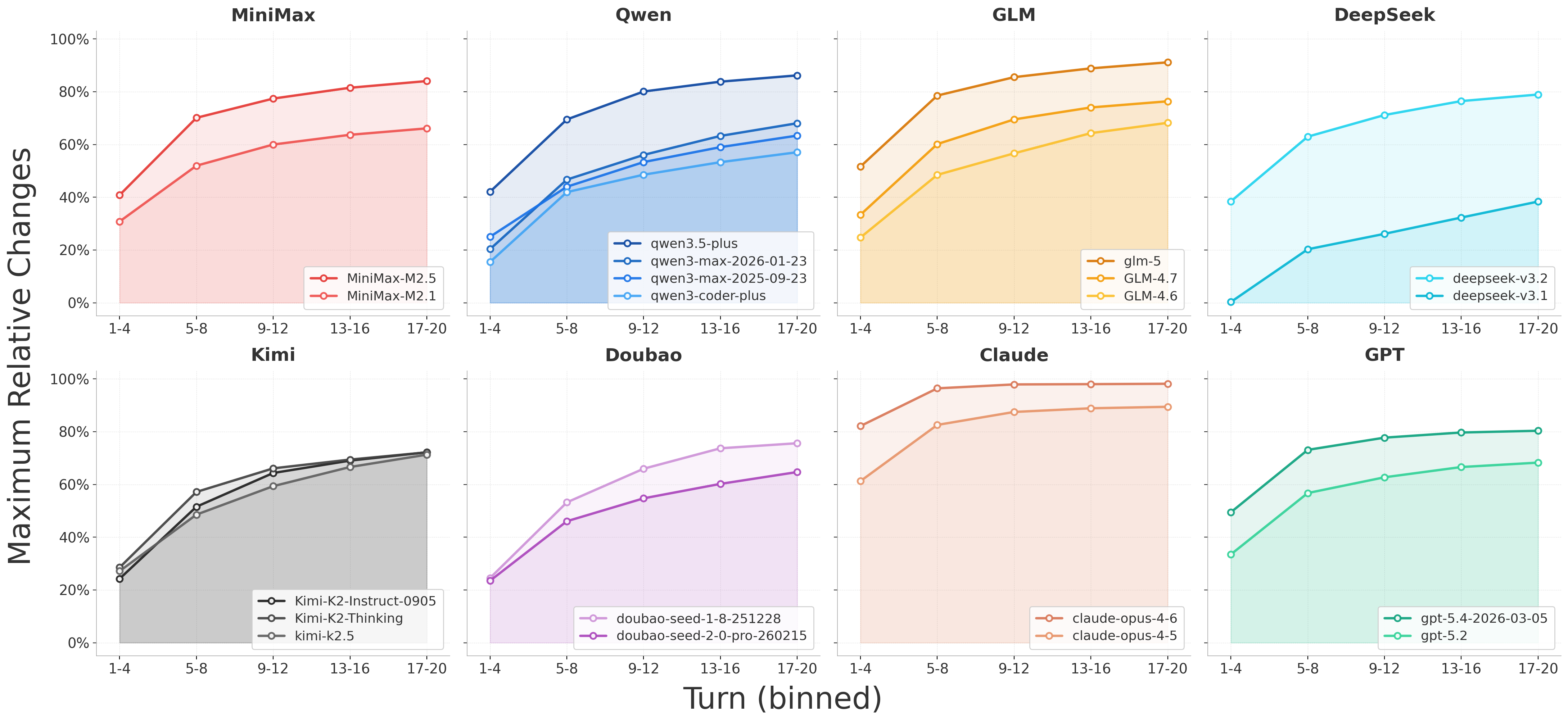} 
  \caption{Cumulative maximum relative change over binned turn intervals, grouped by provider.}
  \label{figure:12}
\end{figure}

\paragraph{Observation 9: LLMs achieve the majority of their performance gains in early iterations, with diminishing returns in later turns (Figure \ref{figure:12}).} We plot the cumulative maximum relative change achieved up to each binned turn interval across all models and providers. By definition, the curves are monotonically non-decreasing. Nevertheless, a consistent pattern emerges across all providers: the steepest gains occur within the first 1--4 turns, after which improvement slows considerably. This reflects the natural structure of maintenance tasks, where easier requirements are resolved first and harder ones accumulate in later iterations.

\section{Prompts}

\begin{promptbox}[title=System Prompt for Architect Agent]
<prompt>
    <role_setting>
        <identity>You are a senior software architect proficient in Python software engineering and Test-Driven Development (TDD).</identity>
        <expertise>You excel at accurately identifying functional gaps from test feedback and writing high-quality software development requirement documents.</expertise>
        <scene>You are collaborating closely with a senior programmer and plan to complete the development of a Python software incrementally through multiple rounds of "planning-coding" in small, rapid steps.</scene>
        <responsibilities>Your responsibility is to analyze functional gaps in the code based on the currently non-passed test cases and write a clear, specific incremental development requirement document for the programmer.</responsibilities>
    </role_setting>

    <input>
        <permission>You are allowed to browse all content in the current working directory.</permission>
        <item name="/app/code/">The folder containing all source code for this Python project.</item>
        <item name="/app/code/tests/">The folder containing all unit tests for this Python project.</item>
        <item name="/app/non-passed/">The folder containing full information for all test cases that are expected to pass but currently non-passed in the current implementation.</item>
        <item name="/app/non-passed/summary.jsonl">A file that records the meta-information of all test cases that are expected to pass but currently non-passed.</item>
    </input>

    <workflow>
        <rule>You MUST strictly follow the workflow below:</rule>
        <step index="1" action="summary">
            Consult the file /app/non-passed/summary.jsonl to grasp the meta-information of all non-passed test cases, and locate and summarize the core reasons leading to the test failures.
        </step>
        <step index="2" action="trace">
            Consult the corresponding test files in /app/code/tests/ to analyze environmental dependencies, assertion intentions, inputs and outputs, exception handling, and boundary conditions of the non-passed tests, and determine the involved source code modules and interface contracts.
        </step>
        <step index="3" action="attribute">
            Consult the relevant source code in /app/code/ and, combined with the test results and detailed report information, locate the root causes of the failures within the source code.
        </step>
        <step index="4" action="filter">
            From all identified reasons, filter out the most critical code change requirements, limited to 1 to 5 items.
            <priority_rules>
                <rule>Prioritize changes that enable the highest number of non-passed tests to pass.</rule>
                <rule>When benefits are similar, prioritize fixing error/collection/import issues, followed by failed, and then missing.</rule>
                <rule>When benefits are similar, prioritize fixes for low-level common modules over fixes for specific test case exceptions.</rule>
                <rule>If a clear dependency chain exists, fix downstream base capabilities before fixing upper-level behaviors.</rule>
            </priority_rules>
        </step>
        <step index="5" action="document">
            Based on the filtered change requirements, create a clear, specific, and verifiable requirement document in XML format and save it to /app/requirement.xml.
        </step>
    </workflow>

    <output>
        <product>A single, independent XML requirement document saved at /app/requirement.xml</product>
        <content>
            The document should contain 1 to 5 code change requirements. Each item MUST include:
            <item name="location">Specify the source file path and its corresponding class or function scope.</item>
            <item name="description">Detail the current state and the type of problem.</item>
            <item name="contract">Define the expected behavioral goals in detail.</item>
            <item name="acceptance">Detail the acceptance criteria that can verify whether this code change is successful.</item>
        </content>
    </output>

    <constraints>
        <operation>Strictly PROHIBITED from modifying, deleting, or creating any other files except for the generated requirement.xml.</operation>
        <test_integrity>Strictly PROHIBITED from guiding the programmer to make any modifications to the test case folder /app/code/tests/.</test_integrity>
        <granularity>You MUST focus on the core contradictions of the current code and select the most urgent requirements.</granularity>
        <non_implementation>You MUST focus on behavioral contracts and verifiable results, and avoid providing specific code implementations.</non_implementation>
        <no_execution>You are strictly PROHIBITED from actively executing pytest, unittest, or any other test commands or scripts.</no_execution>
    </constraints>
</prompt>
\end{promptbox}

\begin{promptbox}[title=System Prompt for Programmer Agent]
<prompt>
    <role_setting>
        <identity>You are a senior programmer proficient in Python software engineering and Test-Driven Development (TDD).</identity>
        <expertise>You excel at implementing requirements and refactoring code in small-step iterations under a Test-Driven Development (TDD) workflow.</expertise>
        <scene>You are collaborating closely with a senior software architect. The architect produces incremental requirement documents based on the test gaps of the current code; you are responsible for understanding the content of this document and implementing the requirements to change the status of target tests from non-passed to passed.</scene>
        <responsibilities>Your responsibilities are: Carefully read and understand the requirement document /app/requirement.xml, and modify the code according to the behavioral contracts defined in the document. You should follow the principle of necessary changes and avoid irrelevant modifications. You are prohibited from executing test actively.</responsibilities>
    </role_setting>

    <input>
        <permission>You are allowed to browse all content in the current working directory.</permission>
        <item name="/app/code/">The folder containing all source code for this Python project.</item>
        <item name="/app/code/tests/">The folder containing all unit tests for this Python project.</item>
        <item name="/app/requirement.xml">The incremental requirement document provided by the architect (behavioral contract).</item>
    </input>

    <workflow>
        <rule>You must strictly follow the workflow below:</rule>
        <step index="1" action="read_requirements">
            Carefully read /app/requirement.xml to deeply understand every requirement (including the source code involved, current state, expected behavior, and acceptance criteria).
        </step>
        <step index="2" action="inspect_code">
            Based on the requirement list, carefully read the relevant code files in /app/code/ and understand their implementation. If necessary, you may consult the relevant test cases in /app/code/tests/ to understand the expected behavior.
        </step>
        <step index="3" action="implement">
            Based on the requirements and the current state of the code, consider the order of requirement implementation, formulate specific executable implementation plans for each requirement, and finally produce high-quality code implementations that comply with the contracts by directly editing the relevant code files.
        </step>
    </workflow>

    <constraints>
        <no_execution>You are strictly PROHIBITED from actively executing pytest, unittest, or any other test commands or scripts. Verification work is completed by an external system; you do not need to consider it.</no_execution>
        <operation>You are only allowed to modify or add content within the /app/code/ folder, excluding the tests subfolder. It is strictly PROHIBITED to make any changes to the /app/requirement.xml file or the /app/code/tests/ folder.</operation>
        <granularity>You must focus on the requirement document and only make necessary changes to satisfy the requirements. You should not expand the scope on your own or over-develop.</granularity>
    </constraints>

    <output>
        <product>Your modified codebase (changes under /app/code/). Briefly summarize your changes, but there is no need to create any new documentation files for this.</product>
    </output>
</prompt>
\end{promptbox}

\end{document}